\documentclass[11pt]{article}
\usepackage{latexsym}  
\usepackage{amssymb}
\usepackage{graphicx}
\usepackage{amsmath}
\numberwithin{equation}{section}

\newcommand{\be}{\begin{equation}}
\newcommand{\ee}{\end{equation}}
\newcommand{\bea}{\begin{eqnarray}}
\newcommand{\eea}{\end{eqnarray}}
\newcommand{\bref}[1]{(\ref{#1})}
\begin{document}
\begin{titlepage}
\begin{flushright}
\end{flushright}
\vspace{4\baselineskip}
\begin{center}
{\Large\bf  Determination of the unknown absolute neutrino mass and MNS parameters at the LHC in the Higgs triplet model.}
\end{center}
\vspace{1cm}
\begin{center}
{\large Hiroyuki Nishiura$^{a,}$
\footnote{E-mail:nishiura@is.oit.ac.jp}}
and
{\large Takeshi Fukuyama$^{b,}$
\footnote{E-mail:fukuyama@se.ritsumei.ac.jp}}
\end{center}
\vspace{0.2cm}
\begin{center} 
${}^{a} $ {\small \it Faculty of Information Science and Technology, 
Osaka Institute of Technology,\\ Hirakata, Osaka 573-0196, Japan}\\[.2cm]
${}^{b}$ {\small \it Department of Physics and R-GIRO, Ritsumeikan University,
Kusatsu, Shiga, 525-8577, Japan}
\medskip
\vskip 10mm
\end{center}
\vskip 10mm
\begin{abstract}
Assuming the Higgs triplet model, we obtain the bounds on the absolute neutrino mass 
and the unknown MNS parameters by measuring decay processes of doubly charged particles at the CERN LHC.
Majorana CP violating phases affect the prediction rather seriously, which is served to restrict these phases.
\end{abstract}
\vspace{3mm}


\end{titlepage}
\section{Introduction}

Presently, observation of the Majorana nature of neutrinos and 
the absolute value of the neutrino mass is indispensable 
for constructing a concrete model for particle physics beyond the standard model (SM).
For instance, renormalizable minimal SO(10) GUT predicts all of the parameters 
of the Maki-Nakagawa-Sakata (MNS) lepton mixing matrix \cite{Maki} including the Dirac and Majorana phases 
unambiguously which plays an essential role to match up with the neutrino oscillation data \cite{Fukuyama1}. 
However, these phases are still left unknown and the experimental observation of them is 
an important issue for near future experiments. 
The GUT model is the comprehensive theory and must explain the whole range of particle physics.  
On the other hand, the Higgs triplet model (HTM) \cite{HTM} is the simplest extension 
of the SM which invokes the new phenomena of lepton physics. 
The HTM may be rather restrictive and may be interpreted as an effective theory. 
However, it makes us easier to extract an essential point of new physics if it works well. 
The HTM is also very interesting because it gives sizable effects on low energy lepton physics.

In \cite{Fukuyama} we showed that the HTM enables us to detect the Majorana property 
by the precise measurement of the usual muon decay.
The interference terms in muon decay due to the Majorana property were first discussed in \cite{Kotani},  
but the detection of them was far beyond the present upper bound.
On the contrary, the HTM gives a rather marginal value to the present precision order.
In the literatures \cite{Han} it was discussed that the unknown MNS parameters may be solved by the observations at the CERN Large Hadron Collider (LHC) if the HTM works well. 
In the previous paper \cite{N-F} we published a brief report on determining 
the lower bound of the minimal neutrino mass, equivalently, the absolute neutrino masses 
by measuring the ratio of decay widths of doubly charged Higgs boson to leptonic channels,
$\frac{\Gamma(\Delta^{--}\rightarrow ee)}{\Gamma(\Delta^{--}\rightarrow \mu\mu)}$, at the LHC. 
There it was shown that the Majorana phases play an essential role allowing this ratio to be both larger and smaller than 1.
In this paper we develop the arguments in more detail and more general than\ \cite{N-F}.

In Sec.~2, we briefly explain the HTM and experimental situations on the considering process at the LHC.
The lepton mixing matrix and the neutrino mass hierarchy are examined in Sec.~3.
Numerical calculations are presented in Sec.~4.   
Finally, Sec.~5 is devoted to discussions.

\section{The Higgs triplet model}
The neutrino-Higgs coupling in the HTM is given by
\begin{align}
{\mathcal L}_\text{HTM}^{} &= \overline{L^c}h_M\,i\tau_2\Delta
L+\text{H. c.}
\end{align}
Here $L\equiv (\nu_L,l_L)^T$ and $L^c\equiv C\overline{L}^T$, and neutrinos are required to be Majorana particles. 
The symmetric $3\times 3$ matrix
$(h_M)_{ll'}~(l,l'=e,\mu,\tau)$ is the coupling strength and $\tau_i(i=1,2,3)$ denote 
the Pauli matrices.
The triplet Higgs boson field with hypercharge $Y=2$ 
can be parameterized by
\begin{align}
\Delta=\begin{pmatrix}\Delta^+/\sqrt2&\Delta^{++}\\
\frac{v_\Delta^{}}{\sqrt2}+\Delta^0&-\Delta^+/\sqrt2\end{pmatrix},
\end{align}
where $v_\Delta^{}$ is the vacuum expectation value of 
the triplet Higgs boson. 
Mass eigenvalues of neutrinos are determined by diagonalization of
$m_\nu^{}=\sqrt2h_M^{}v_\Delta^{}$. 
There is a tree level contribution to the electroweak $\rho$ parameter 
from the triplet vacuum expectation value as 
$\rho\thickapprox 1-2v_\Delta^2/v^2$.
The CERN LEP precision results can give an upper limit $v_\Delta^{}\lesssim 5$ GeV. 
There is no stringent bound from the quark sector on triplet Higgs bosons 
because they do not couple to quarks. 

The Yukawa interaction of the singly and the doubly charged Higgs bosons
is written as
\begin{align}
{\mathcal L}_{\Delta}^{} =& 
-\sqrt2(h_M^\dag U)_{\ell i}\overline{\ell_L^{}}N_i^c\Delta^-\nonumber \\
&-(h_M^\dag)_{\ell\ell'}\overline{\ell_L^{}}{{\ell'}_L}^c\Delta^{--}
+\text{H.c.},
\end{align}
where 
\be
h_M^{}=Um_\nu^\text{diag}U^T/(\sqrt2v_\Delta^{})\equiv \left<m_\nu\right>_{ab}/(\sqrt2v_\Delta^{}), 
\ee
and 
$N_i(i=1,2,3)$ represent Majorana neutrinos which satisfy the conditions $N_i=N_i^c=C\overline{N_i}^T$. 
The most stringent constraint on the triplet Yukawa coupling 
comes from $\mu\to ee{\bar e}$ through the tree-level contribution due to 
the doubly charged Higgs boson~\cite{Mu3eExp}. 
Thus the peculiar properties of the HTM appear in the processes of the doubly charged Higgs.
Among them, we have a sizable cross section of 
$\Delta^{++}\rightarrow l_al_b$ for $m_\Delta=O(100)$ GeV,
and the decay width of the doubly charged Higgs boson to this leptonic channel is given by
\be
\Gamma(\Delta^{++}\rightarrow l_a^+l_b^+)=\frac{1}{4\pi(1+f)}|h_{ab}|^2m_{\Delta^{++}},
\label{ratio}
\ee
where $f=1(0)$ for $a=b$ ($a\neq b$). 

Searching for the neutrino mass in the HTM at the LHC was discussed in \cite{Han},
where the event numbers were estimated at the scheduled energy and luminosity.
Unfortunately, the LHC was forced to lower the energy scale $10$ Tev and the luminosity  $10^{33}$ cm$^{-2}$s$^{-1}$.
The cross section for $pp\rightarrow \Delta^{++}\Delta^{--}$, 
\bea
  99 &&\mbox{fb  for}~ m_{\Delta^{++}} = 200\mbox{GeV},\nonumber\\
 5.9 &&\mbox{fb  for}~ m_{\Delta^{++}} = 400\mbox{GeV},
\label{fb1}
\eea
for $\sqrt {s} = 14$TeV is reduced to
\bea
  53 &&\mbox{fb  for}~ m_{\Delta^{++}} = 200\mbox{GeV},\nonumber\\
 2.6 &&\mbox{fb  for}~ m_{\Delta^{++}} = 400\mbox{GeV},
\label{fb2}
\eea
for $\sqrt{s} = 10$ TeV~\cite{N-F}.

The first 100-day run at low luminosity resulted the integrated luminosity $10$ fb$^{-1}$.
So we may have a sizable number of events even in this case, 
though the final number of $ll$ events depends on the Br($\Delta\rightarrow ll$). 
The decay width to the other channel, WW channel is 
\be
\Gamma(\Delta^{--}\rightarrow W^-W^-)\approx \frac{v_\Delta^2m_{\Delta^{++}}^3}{2\pi v^4}\equiv cv_\Delta^2,
\label{WW}
\ee
(full expression is given in \cite{Han2}) and the branching ratio is given by
\be
Br(\Delta\rightarrow l_al_b)\equiv Br_{ab}=\frac{|\left<m_\nu\right>_{ab}|^2}{\Sigma_{a\geq b}|\left<m_\nu\right>_{ab}|^2+4c^2v_\Delta^4}.
\ee
Here we have neglected $\Delta^{--}\rightarrow W^-\Delta^-,~\Delta^{--}\rightarrow \Delta^-\Delta^-$ since we assumed $m_{\Delta^{--}}\approx m_{\Delta^-}$.
Otherwise the Higgs triplet gives too large loop correction to $\rho$ parameter \cite{rhoHTM}.
The estimate of \bref{ratio} and \bref{WW} indicates that \bref{ratio} is the dominant process.
In order to circumvent the ambiguous factors $m_\Delta$ and $v_\Delta$, let us consider the ratios
\be
\frac{\Gamma(\Delta^{--}\rightarrow ee)}{\Gamma(\Delta^{--}\rightarrow \mu\mu)}
=\left|\frac{\left<m_\nu\right>_{ee}}{\left<m_\nu\right>_{\mu\mu}}\right|^2,
\ee
\be
\frac{\Gamma(\Delta^{--}\rightarrow \mu e)}{\Gamma(\Delta^{--}\rightarrow \mu\mu)}
=\left|\frac{\left<m_\nu\right>_{\mu e}}{\left<m_\nu\right>_{\mu\mu}}\right|^2.
\ee
Here the averaged masses $\left<m_\nu\right>_{a b}\ (a,b =e, \mu, \tau)$ are defined by 
\bea
\left<m_\nu\right>_{a b}&=&U_{a 1}U_{b 1}m_1+U_{a 2}U_{b 2} m_2+U_{a 3}U_{b 3} m_3,
\eea
where $U_{a i} \ (i=1,2,3)$ are the components of the MNS lepton mixing matrix 
and $m_i\ (i=1,2,3)$ are the neutrino masses. 
We have defined the averaged mass without an absolute symbol unlike the conventional one 
since $h_M$ is symmetric and complex matrix in general.

\section{Lepton Mixing matrix and neutrino mass hierarchy}
We know that the MNS lepton mixing matrix $U$ is well approximated 
by the tribimaximal matrix \cite{tri},

\be
U=
\left(
        \begin{array}{ccc}
         \sqrt{\frac{2}{3}}  & \sqrt{\frac{1}{3}}e^{i\beta} &  0 \\
        -\sqrt{\frac{1}{6}}e^{-i\beta} & \sqrt{\frac{1}{3}} & -\sqrt{\frac{1}{2}}e^{i(\rho -\beta)}\\
         -\sqrt{\frac{1}{6}}e^{-i\rho}  & \sqrt{\frac{1}{3}}e^{-i(\rho-\beta)} & \sqrt{\frac{1}{2}}
        \end{array}
\right),
\label{TBM}
\ee
which is supplemented with the Majorana phases, $\beta$ and $\rho$.
If we neglect the Majorana phases, it is a special case of the $\mu-\tau$ symmetric model \cite{F-N}, 
\begin{equation}
U=\left(
\begin{array}{ccc}
c_1&s_1&0\\
-\frac{1}{\sqrt{2}}s_1&\frac{1}{\sqrt{2}}c_1&-\frac{1}{\sqrt{2}}\\
-\frac{1}{\sqrt{2}}s_1&\frac{1}{\sqrt{2}}c_1&\frac{1}{\sqrt{2}}
\end{array}
\right),
\label{KM}
\end{equation}
with  sin$\theta_{12}\equiv s_1=\sqrt{\frac{1}{3}}$ and cos$\theta_{12}\equiv c_1=\sqrt{\frac{2}{3}}$.
Equation \bref{KM} satisfies the $\mu-\tau$ symmetry in more general form than \bref{TBM}.

We have assumed so far the tribimaximal approximation \bref{TBM} with a vanishing (1,3) element, 
that is $\theta_{13}=0$. 
In this case we have 
\be
\text{Br}_{e\mu}=\text{Br}_{e\tau},~\text{Br}_{\mu\mu}=\text{Br}_{\tau\tau}, 
\label{mutau}
\ee
irrespectively of $\theta_{12}$.
The Majorana phases with $\beta\neq \rho$ break the $\mu-\tau$ symmetry, 
but the above relations still hold since the corresponding averaged masses are equal up to the overall phase.
It should be noticed that we have no approximation for the Majorana phases but
have the small effects of $\theta_{13}$ and, therefore, of the Dirac phase.
Also we may have some small deviation from sin$\theta_{12}=\sqrt{\frac{1}{3}}$.

In this paper, we first approximate the MNS lepton matrix as \bref{TBM}, and  proceed to extend it to the $\theta_{13}\neq 0$ case.
We obtain a generalized lepton mixing matrix which tends to the tribimaximal one in the limit of $\theta_{13}=0$,
\be
U=
\left(
        \begin{array}{ccc}
         c\sqrt{\frac{2}{3}}  & c\sqrt{\frac{1}{3}}e^{i\beta} &  se^{i(\rho-\delta)} \\
        \left(-\sqrt{\frac{1}{6}}+s\sqrt{\frac{1}{3}}e^{i\delta}\right)e^{-i\beta} & \left(\sqrt{\frac{1}{3}}+ s\sqrt{\frac{1}{6}}e^{i\delta}\right)&-c\sqrt{\frac{1}{2}}e^{i(\rho -\beta)}\\
         \left(-\sqrt{\frac{1}{6}}-s\sqrt{\frac{1}{3}}e^{i\delta}\right)e^{-i\rho}  & \left(\sqrt{\frac{1}{3}}-s\sqrt{\frac{1}{6}}e^{i\delta}\right)e^{-i(\rho-\beta)} & c\sqrt{\frac{1}{2}}
        \end{array}
\right),
\ee
where $s \equiv \mbox{sin}\theta_{13}$, $c \equiv  \mbox{cos}\theta_{13}$, 
and the $\delta$ is the CP violating Dirac Phase.

In numerical analysis, we use experimental values of neutrino mass squared differences~\cite{PDG},
$\Delta m_{sol}^2$ and $\Delta m_{atm}^2$ with assuming the normal and the inverse hierarchy for neutrino masses:

For the normal hierarchy (NH) case, the neutrino masses are given by
\be
m_1=m_0,~~
m_2=\sqrt{m_0^2 + \Delta m_{sol}^2},~~
m_3=\sqrt{m_0^2 + \Delta m_{sol}^2 + \Delta m_{atm}^2},
\ee
where the smallest neutrino mass is denoted as $m_0$.

For the inverse hierarchy (IH) case, the neutrino masses are given by
\be
m_3=m_0,~~
m_2=\sqrt{m_0^2+\Delta m_{atm}^2},~~
m_1=\sqrt{m_0^2+\Delta m_{atm}^2-\Delta m_{sol}^2},
\ee
where the smallest neutrino mass is denoted as $m_0$, as before.

We adopt the center values of the neutrino mass squared differences~\cite{PDG} as
\be
\Delta m_{sol}^2=(8.0\pm 0.3)\times 10^{-5}~\text{eV}^2,~~\Delta m_{atm}^2=(1.9 - 3.0)\times 10^{-3}~\text{eV}^2
\ee

\section{Numerical Calculations}
Let us consider the following ratios for the decay widths of the doubly charged Higgs boson to leptonic channels including $e$ and $\mu$ flavors
\be
\frac{\Gamma(\Delta^{--}\rightarrow ee)}{\Gamma(\Delta^{--}\rightarrow \mu\mu)}
=\left|\frac{\left<m_\nu\right>_{ee}}{\left<m_\nu\right>_{\mu\mu}}\right|^2,
\label{eemumu}
\ee
\be
\frac{\Gamma(\Delta^{--}\rightarrow e\mu)}{\Gamma(\Delta^{--}\rightarrow \mu\mu)}
=\left|\frac{\left<m_\nu\right>_{e\mu}}{\left<m_\nu\right>_{\mu\mu}}\right|^2.
\label{emumumu}
\ee
The ratios defined by Eqs.\bref{eemumu}\ and \bref{emumumu} are functions of three parameters, 
namely, the smallest neutrino mass $m_0$, and the CP violating Majorana phases $\beta$ and $\rho$ 
if the mixing angles and the Dirac phase $\delta$ are fixed.
Therefore, a measurement of the ratio at the LHC will lead us to a constraint among these three parameters.
That is, given the ratio of \bref{eemumu}, and so on, 
the lower bound of $m_0$ is obtained as the minimum value of $m_0$ 
when we make $\beta$ and $\rho$ run over all possible values both in the NH  and the IH cases for the neutrino mass  hierarchy. 
Note that for the tribimaximal mixing case with $\theta_{13}=0$ 
the Dirac phase $\delta$ does not appear. 

In our previous paper \cite{N-F}, we focused only on the ratio of \bref{eemumu} in the tribimaximal mixing case with $\theta_{13}=0$.
This behavior is presented in Fig.~1(a) and Fig.~1(b) for the NH and the IH cases, respectively. 
The nonshaded area in Fig.~1 is allowed, and is obtained by running over all possible values of $\beta$ and $\rho$.
Namely, it is shown \cite{N-F} that the measurement of the ratio at the LHC will fix the lower bound of the smallest neutrino mass $m_0$.
This is an interesting feature of the HTM.
 
In the present paper, we consider effects due to $\theta_{13 } \neq 0$ and other aspects of the model. 
In Fig.~2 and Fig.~3, we present  allowed regions in the plane of the ratio 
$\frac{\Gamma(\Delta^{--}\rightarrow ee)}{\Gamma(\Delta^{--}\rightarrow \mu\mu)}$
and $m_0$ by taking sin$^2\theta_{13}=0.05$ and $\delta=0$ for the NH and IH cases for the neutrino masses. 
The nonshaded area is allowed, and is obtained by running over all possible values of $\beta$ and $\rho$, as before.
We find that effects due to $\theta_{13 } \neq 0$ make the lower bound of the neutrino mass $m_0$ smaller for NH and larger for IH as shown in Fig.~2. 

In Fig.~3, we show an enlarged version of Fig.~2.  It should be noted that the ratio can be either larger or smaller than one.
It is interesting enough that for 
$\frac{\Gamma(\Delta^{--}\rightarrow ee)}{\Gamma(\Delta^{--}\rightarrow \mu\mu)}<0.1$ 
it appears not only the lower bound but also the upper limit of the smallest neutrino mass $m_0$ for the NH case (Fig.~3(a)).
For the IH case (Fig.~3(b)), 
the range over which the lower bound disappears is enlarged relatively to the case of sin$\theta_{13}=0$ of Fig.~2(b).

Considering the other channel $\Delta\rightarrow \mu e$,  we also show the results of \bref{emumumu} in Fig.~4 and Fig.~5.

So far we have considered the constraint on the smallest neutrino mass $m_0$.
The decay ratios of the doubly charged Higgs are also served to determine the other MNS parameters.
For instance, if we fix $m_0$ with zero $\theta_{13}$ and measure the ratios, 
we obtain constraints on the Majorana phases of $\beta$ and $\rho$, some examples of which are shown in Fig.~6. 

One ratio of the decay widths, for instance \bref{eemumu}, gives one constraint among $m_0$, $\beta$, and $\rho$ for the $\theta_{13}=0$ case.
Another different ratio, \bref{emumumu} also gives an independent another constraint.
Therefore if we fix $m_0$ \cite{single}, we can determine the values of  $\beta$ and $\rho$ from the
intersections of the contour curves of two independent ratios, some examples of which are shown in Fig.~7.

\section{Discussions}
We have obtained the lower bound on the smallest neutrino mass $m_0$ 
by considering the decay processes of the doubly charged particle to leptonic channels at the LHC.
In these calculations, we have approximated the MNS matrix by the tribimaximal one supplemented with the Majorana phases.
This approximation has been relaxed to accept nonzero $\theta_{13}$, and we considered its effect on the estimation of $m_0$.

However these considerations are not only restricted on the bound of $m_0$ 
but also on the constraint on the Majorana phases as shown in Fig.~6 and Fig.~7.
Also we may be able to prove the $\mu-\tau$ symmetry by examining \bref{mutau}.

The full MNS parameters are widely analyzed by \cite{MTFN} from
the various experiments such as single beta decay \cite{single}, neutrinoless double beta decay \cite{betadecay} and so on. 
Now the LHC and long baseline experiments \cite{longbase} enter into this survey.
The constraints from these independently different experiments are crucial to the final determination 
of the MNS parameters.

\vspace{0.5cm}

{\bf Acknowledgments}~~~\\[2mm]
We are grateful to H.~Sugiyama, K.~Tsumura, and A.G.~Akeroyd for useful comments.
The work of T.~F. is supported in part by the Grant-in-Aid 
for Scientific Research from the Ministry of Education, 
Science and Culture of Japan (No. 20540282).

\newpage
\begin{center}
{\scalebox{0.8}{\includegraphics{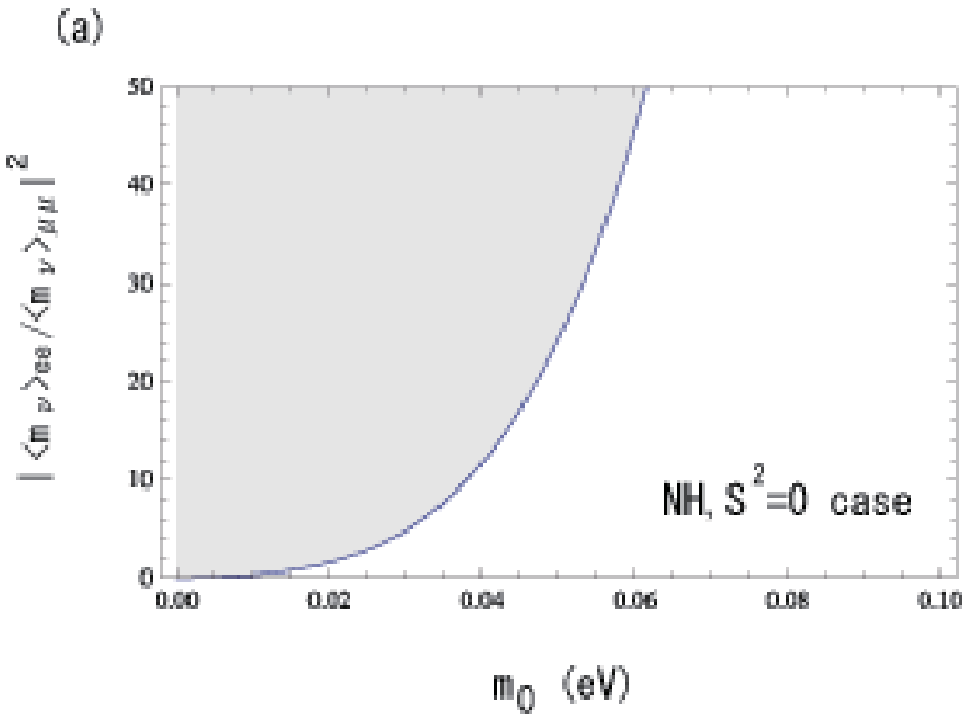}} }
{\scalebox{0.8}{\includegraphics{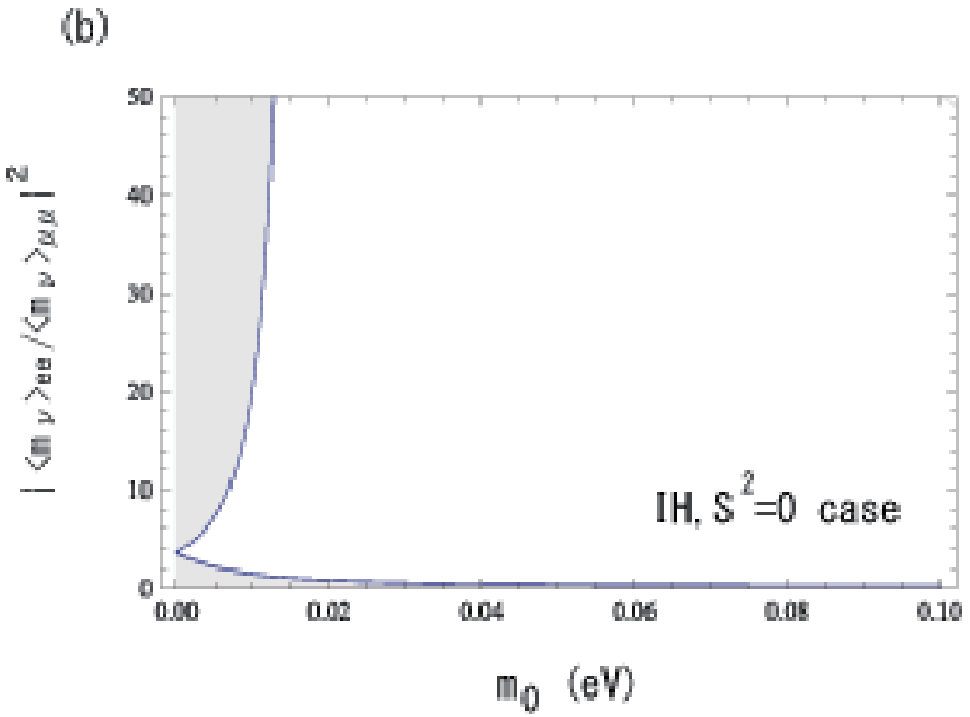}} }
\end{center}
\begin{quotation}
{\bf Fig.~1 }   Behavior of $\frac{\Gamma(\Delta^{--}\rightarrow ee)}{\Gamma(\Delta^{--}\rightarrow \mu\mu)}
=\left|\frac{\left<m_\nu\right>_{ee}}{\left<m_\nu\right>_{\mu\mu}}\right|^2$ 
versus $m_0$ in the case of sin$^2\theta_{13}=0$ cited from \cite{N-F}.
The nonshaded area is allowed, and is obtained by running over all possible values of $\beta$ and $\rho$.
(a) Normal hierarchy case for the neutrino mass with sin$^2\theta_{13}=0$ case. 
(b) Inverse hierarchy case for the neutrino mass with sin$^2\theta_{13}=0$ case.
\end{quotation}

\vspace{1cm}

\begin{center}
{\scalebox{0.8}{\includegraphics{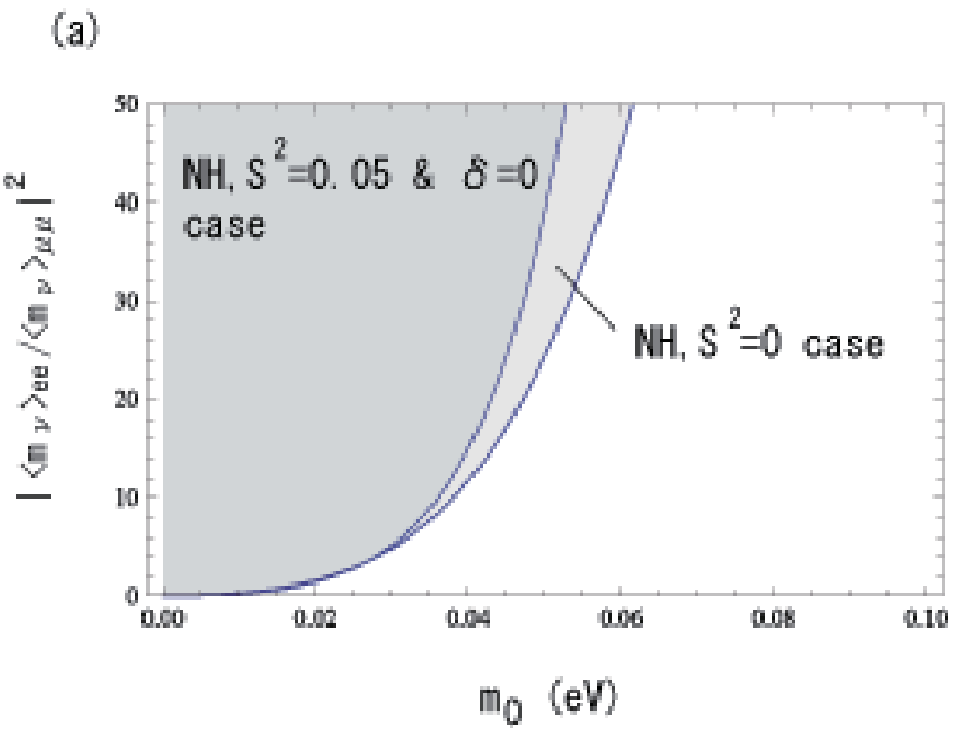}} }
{\scalebox{0.8}{\includegraphics{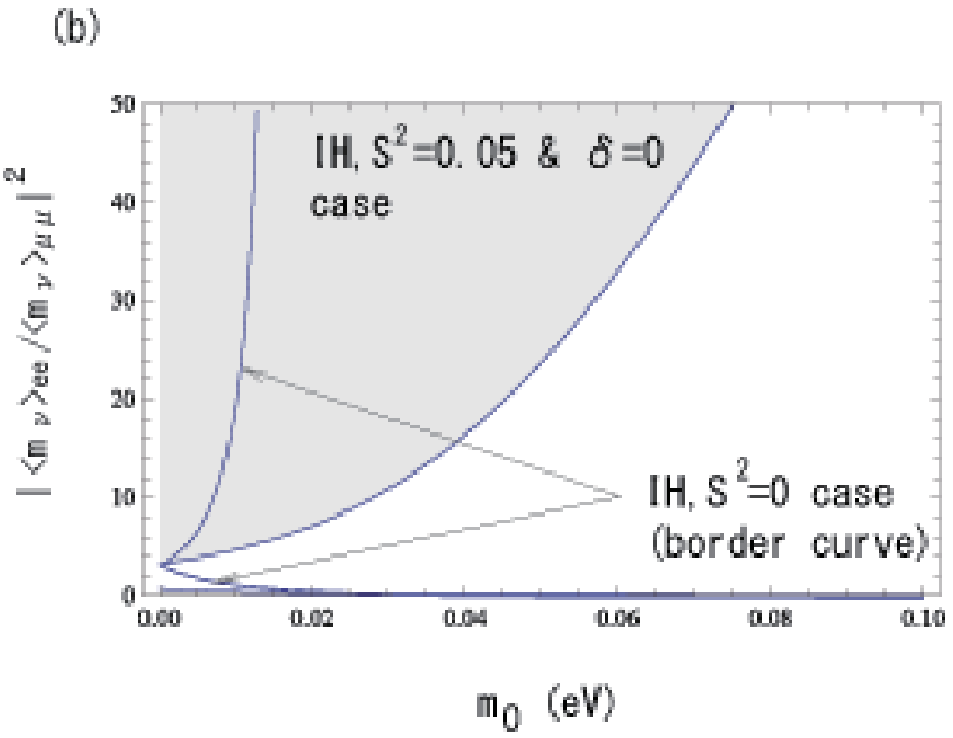}} }
\end{center}
\begin{quotation}
{\bf Fig.~2 }   Behavior of $\frac{\Gamma(\Delta^{--}\rightarrow ee)}{\Gamma(\Delta^{--}\rightarrow \mu\mu)}
=\left|\frac{\left<m_\nu\right>_{ee}}{\left<m_\nu\right>_{\mu\mu}}\right|^2$ 
versus $m_0$ in the case of sin$^2\theta_{13}=0.05$ and $\delta=0$.
The nonshaded area is allowed, and is obtained by running over all possible values of $\beta$ and $\rho$.
(a) and (b) are for the normal and the inverse hierarchy cases, respectively. 
The result of Fig.~1 is overwritten to show the effect of $\theta_{13}$.
\end{quotation}

\begin{center}
{\scalebox{0.8}{\includegraphics{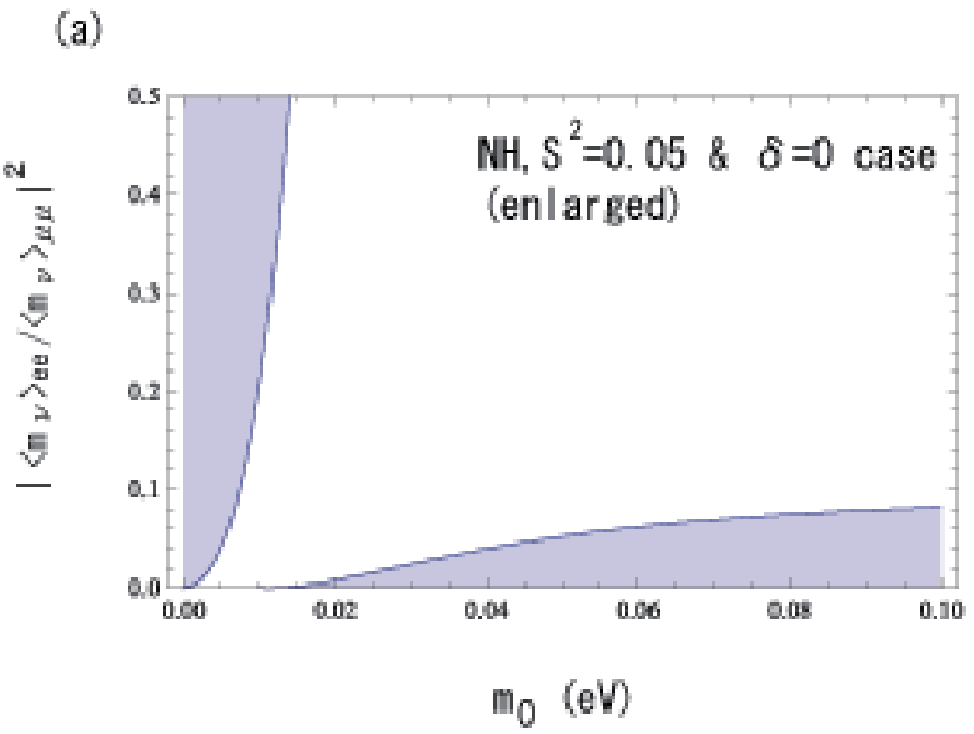}} }
{\scalebox{0.8}{\includegraphics{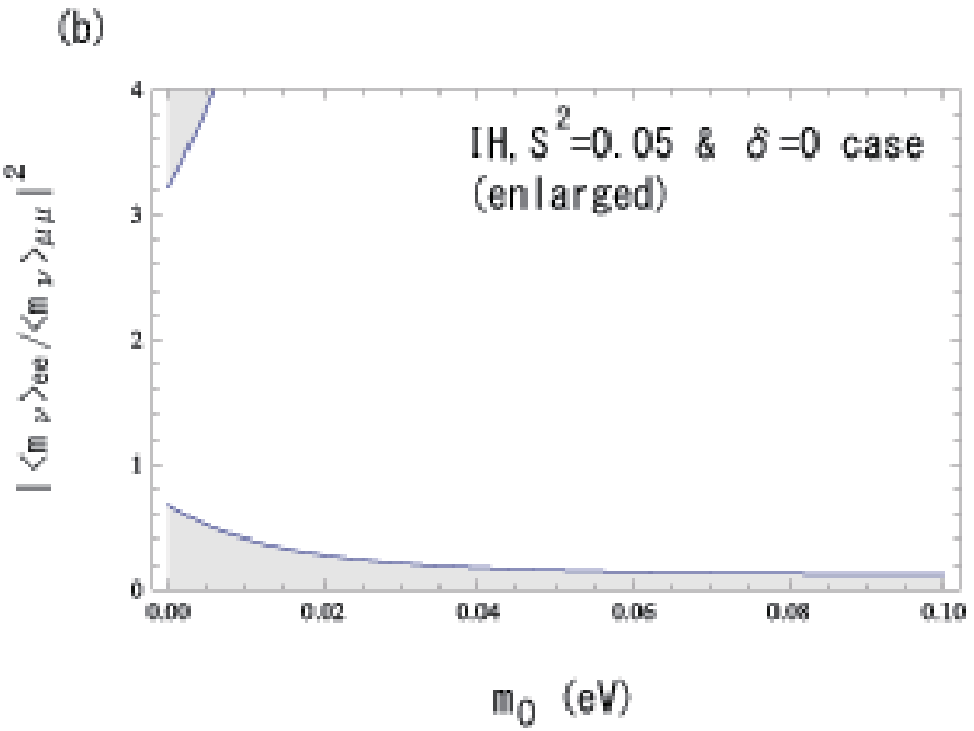}} }
\end{center}
\begin{quotation}
{\bf Fig.~3 }  The same figure as Fig.~2 but with enlarged scales 
for the case of sin$^2\theta_{13}=0.05$ and $\delta=0$.
(a) and (b) are for the normal and the inverse hierarchy cases, respectively.
\end{quotation}

\vspace{1cm}

\begin{center}
{\scalebox{0.8}{\includegraphics{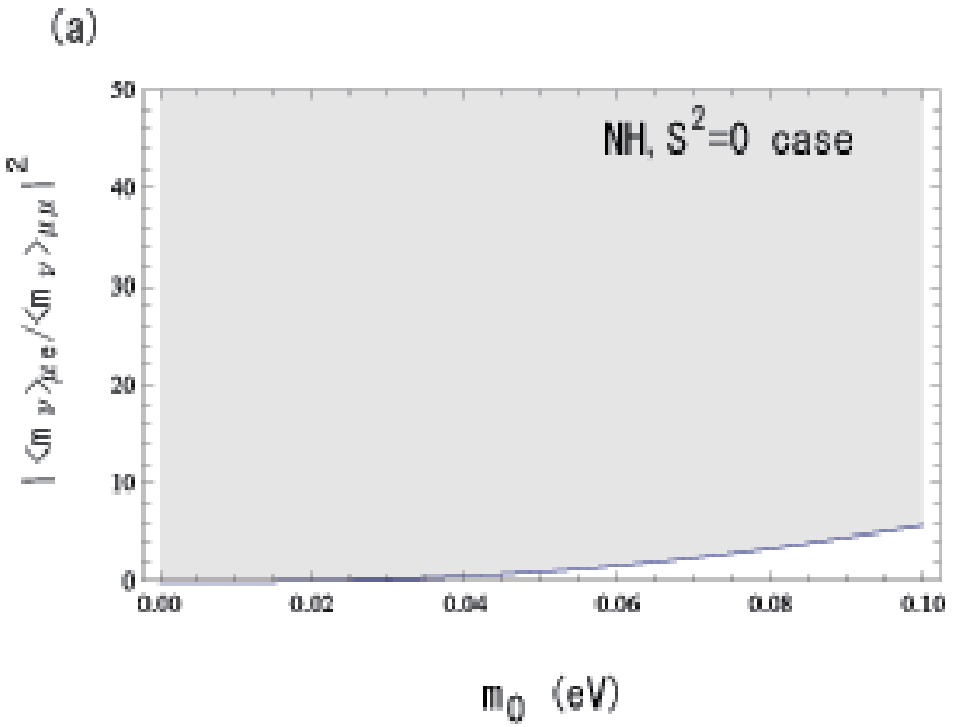}} }
{\scalebox{0.8}{\includegraphics{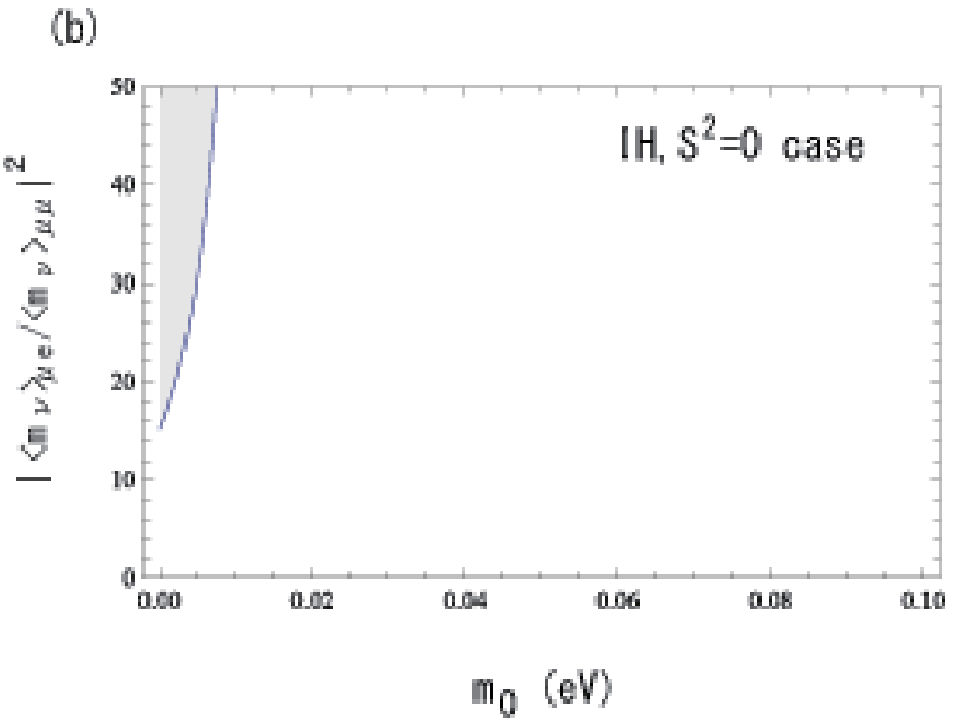}} }
\end{center}
\begin{quotation}
{\bf Fig.~4 }  Behavior of  $\frac{\Gamma(\Delta^{--}\rightarrow \mu e)}{\Gamma(\Delta^{--}\rightarrow \mu\mu)}
=\left|\frac{\left<m_\nu\right>_{\mu e}}{\left<m_\nu\right>_{\mu\mu}}\right|^2$  
versus $m_0$ in the case of sin$^2\theta_{13}=0$.
The nonshaded area is allowed, and is obtained by running over all possible values of $\beta$ and $\rho$.
(a) and (b) are for the normal and the inverse hierarchy cases, respectively.
\end{quotation}

\newpage

\begin{center}
{\scalebox{0.8}{\includegraphics{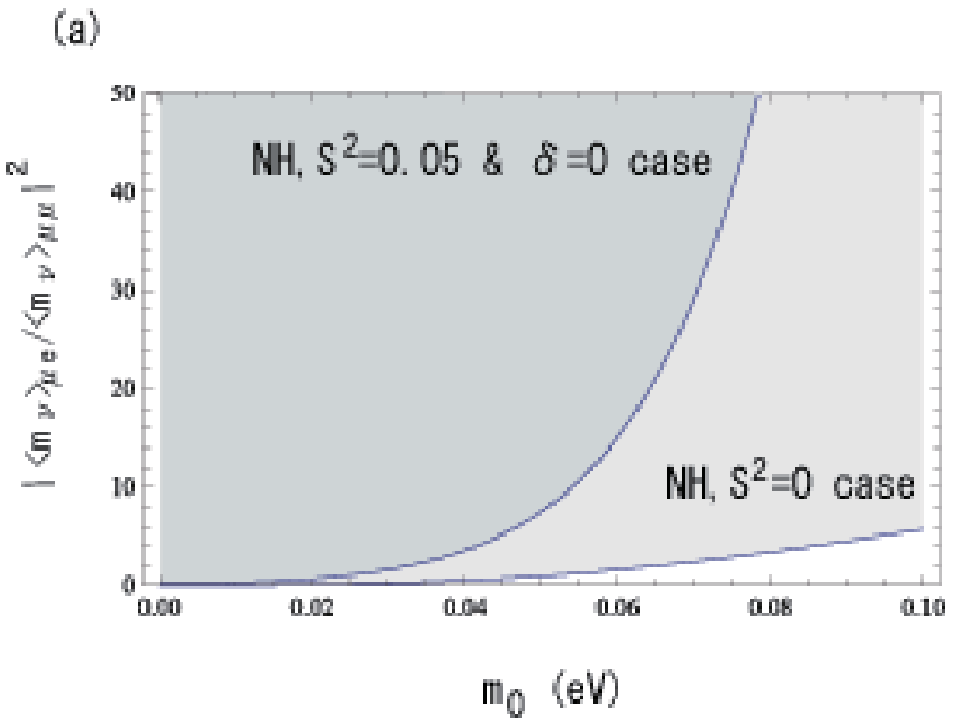}} }
{\scalebox{0.8}{\includegraphics{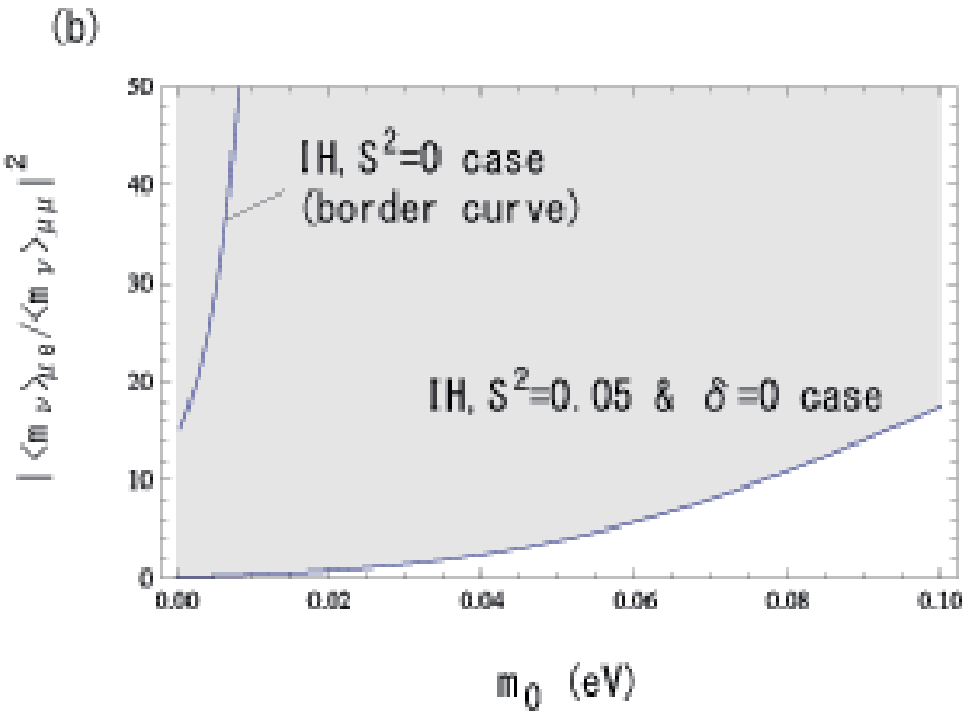}} }
\end{center}
\begin{quotation}
{\bf Fig.~5 }  Behavior of  $\frac{\Gamma(\Delta^{--}\rightarrow \mu e)}{\Gamma(\Delta^{--}\rightarrow \mu\mu)}
=\left|\frac{\left<m_\nu\right>_{\mu e}}{\left<m_\nu\right>_{\mu\mu}}\right|^2$  
versus $m_0$ in the case of sin$^2\theta_{13}=0.05$ and $\delta=0$.
The nonshaded area is allowed, and is obtained by running over all possible values of $\beta$ and $\rho$.
(a) and (b) are for the normal and the inverse hierarchy cases, respectively.
The result of Fig.~4 is overwritten to show the effect of $\theta_{13}$.
\end{quotation}

\newpage

\begin{center}
{\scalebox{0.75}{\includegraphics{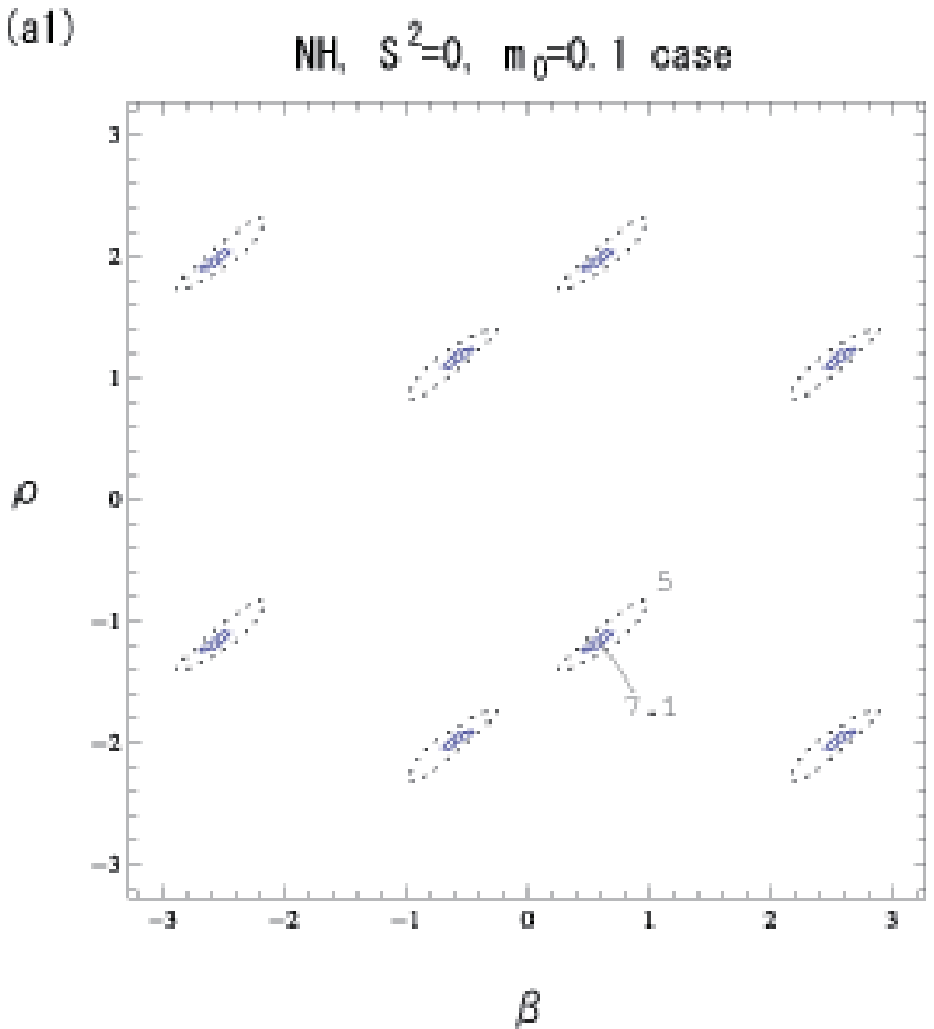}} }\ 
{\scalebox{0.75}{\includegraphics{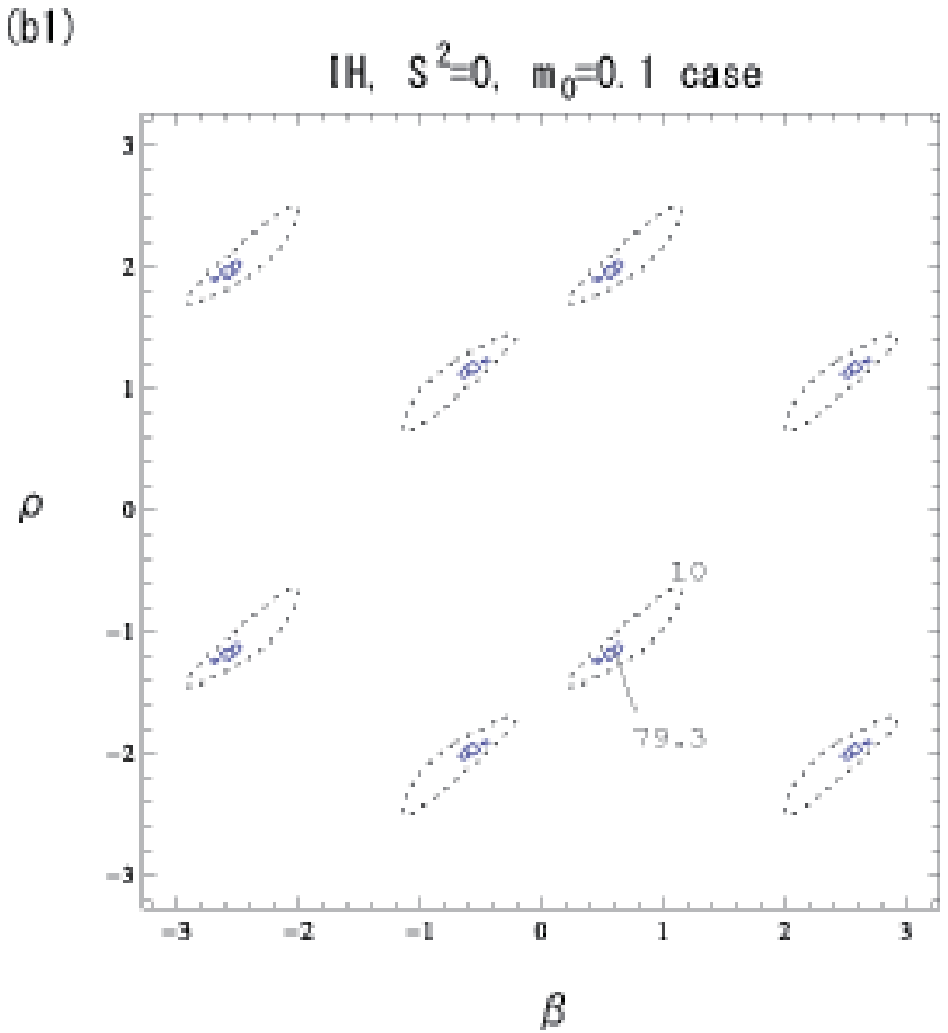}} }\ 
{\scalebox{0.75}{\includegraphics{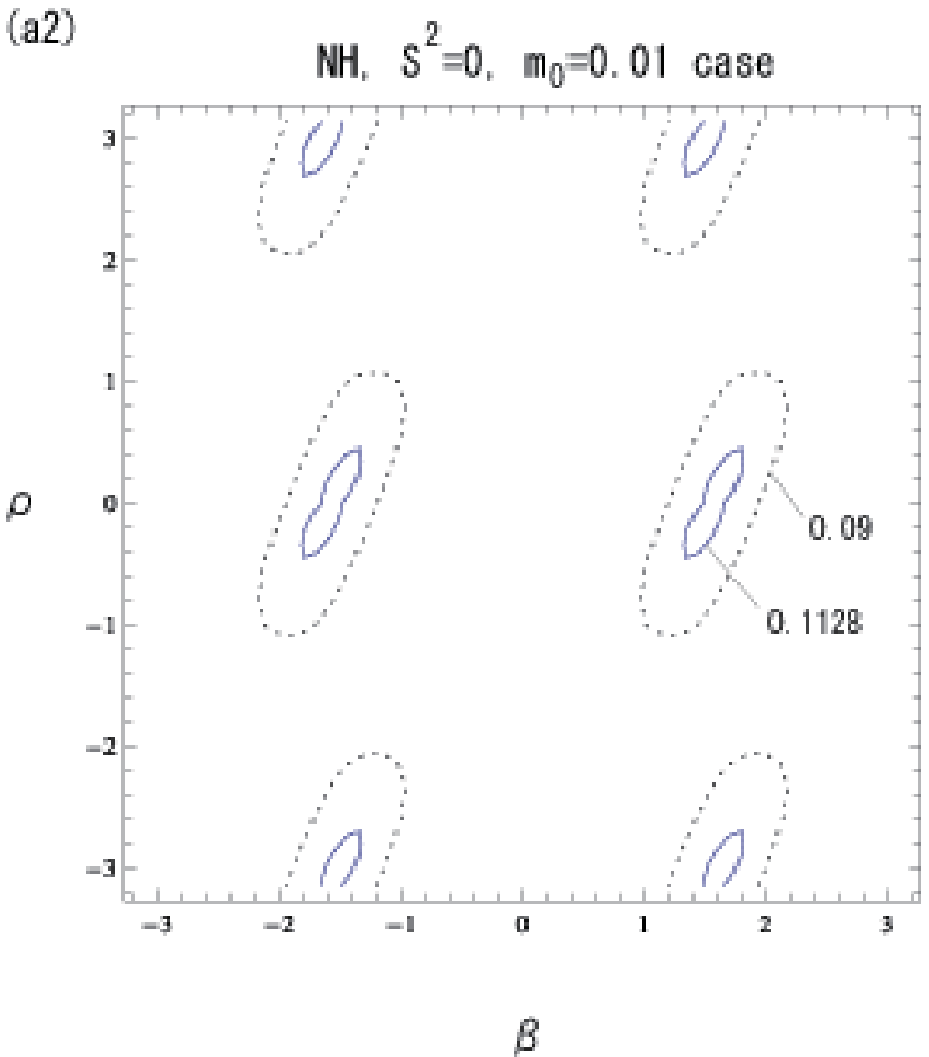}} } \ 
{\scalebox{0.75}{\includegraphics{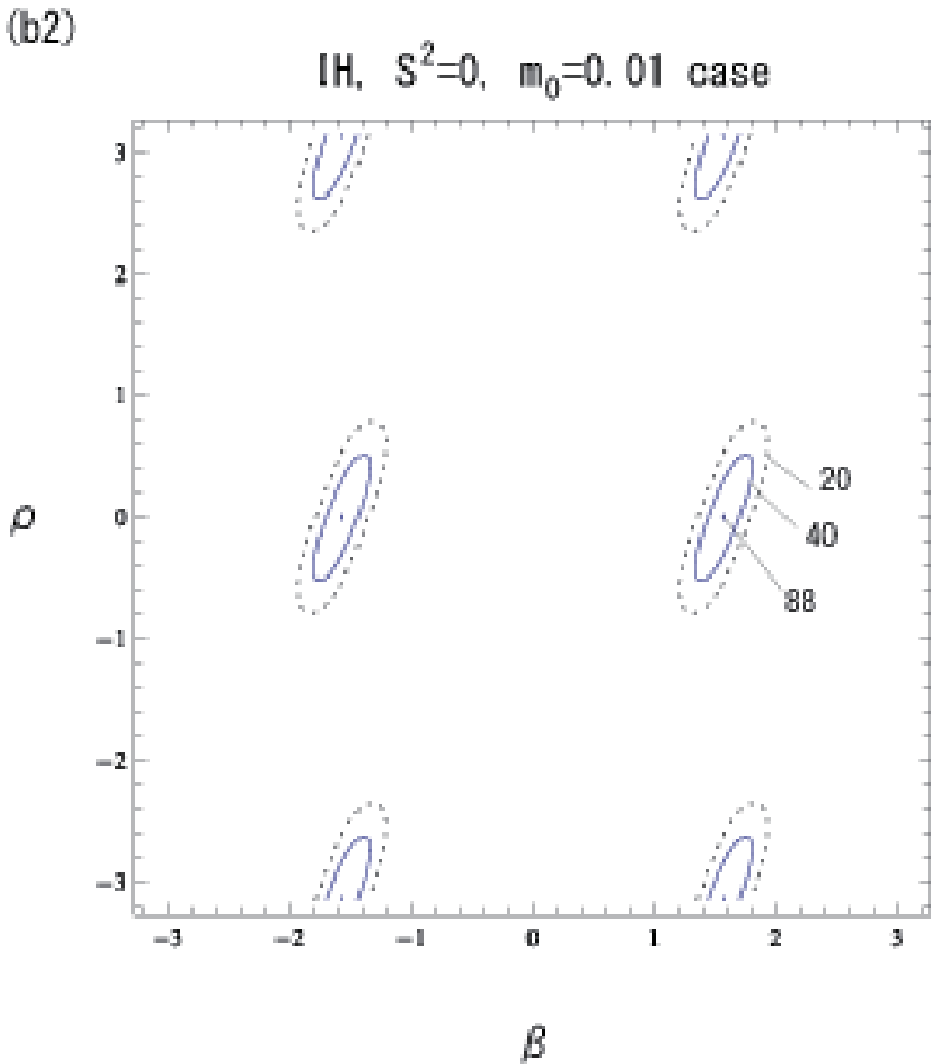}} }
\end{center}
\begin{quotation}
{\bf Fig.~6 }  Contour curves of  $\frac{\Gamma(\Delta^{--}\rightarrow \mu e)}{\Gamma(\Delta^{--}\rightarrow \mu\mu)}$ in the $\beta$ - $\rho$ plane  
 for the case of  $m_0=0.1$ eV (panel 1) and $m_0=0.01$ eV (panel 2) with sin$^2\theta_{13}=0$. 
Indices a and b indicate the normal and the inverse hierarchy cases, respectively. 
For the normal hierarchy cases (a1) and (a2), 
curves with $\frac{\Gamma(\Delta^{--}\rightarrow \mu e)}{\Gamma(\Delta^{--}\rightarrow \mu\mu)}
<7.1$  and  $\frac{\Gamma(\Delta^{--}\rightarrow \mu e)}{\Gamma(\Delta^{--}\rightarrow \mu\mu)}<0.1128$  
are allowed for $m_0=0.1$ eV and  $m_0=0.01$ eV, respectively. For the inverse hierarchy cases (b1) and (b2), 
curves with $\frac{\Gamma(\Delta^{--}\rightarrow \mu e)}{\Gamma(\Delta^{--}\rightarrow \mu\mu)}<79.3$ and 
$\frac{\Gamma(\Delta^{--}\rightarrow \mu e)}{\Gamma(\Delta^{--}\rightarrow \mu\mu)}<88$ are allowed 
for $m_0=0.1$ eV and $m_0=0.01$ eV, respectively.
\end{quotation}

\vspace{1cm}

\newpage

\begin{center}
{\scalebox{0.75}{\includegraphics{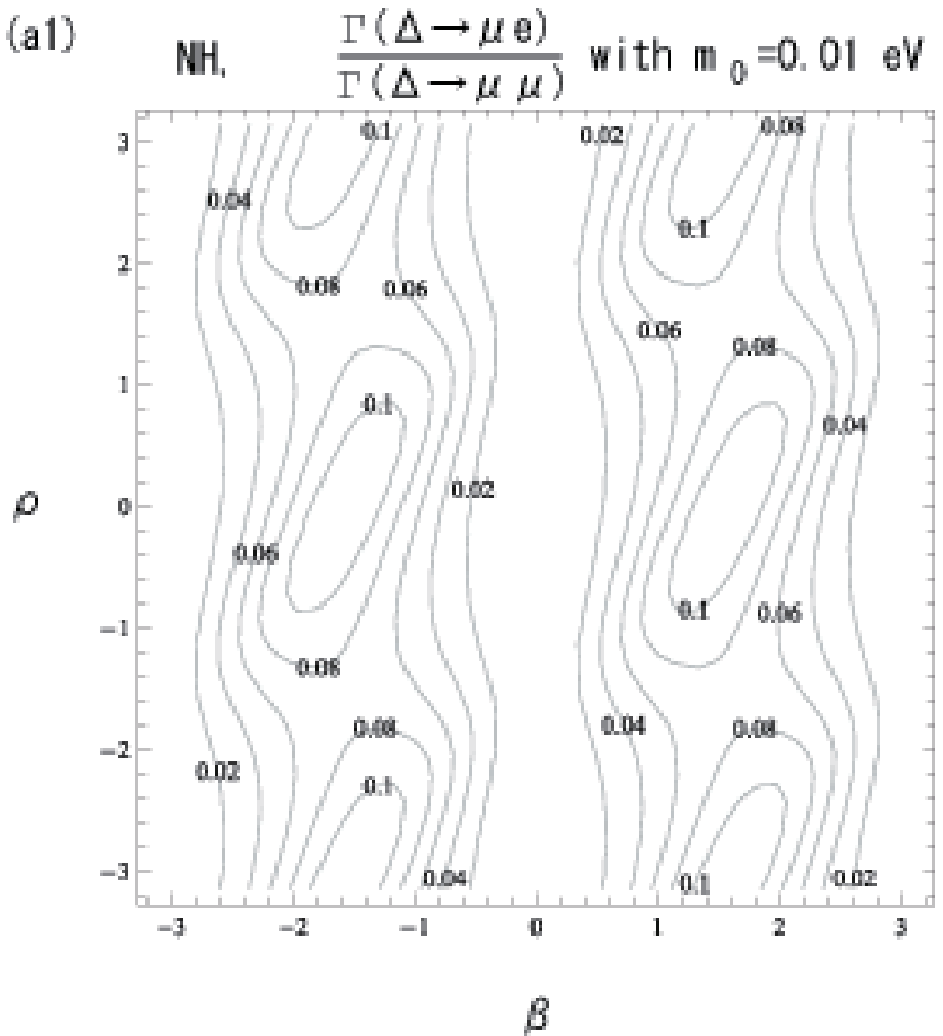}} }\ 
{\scalebox{0.75}{\includegraphics{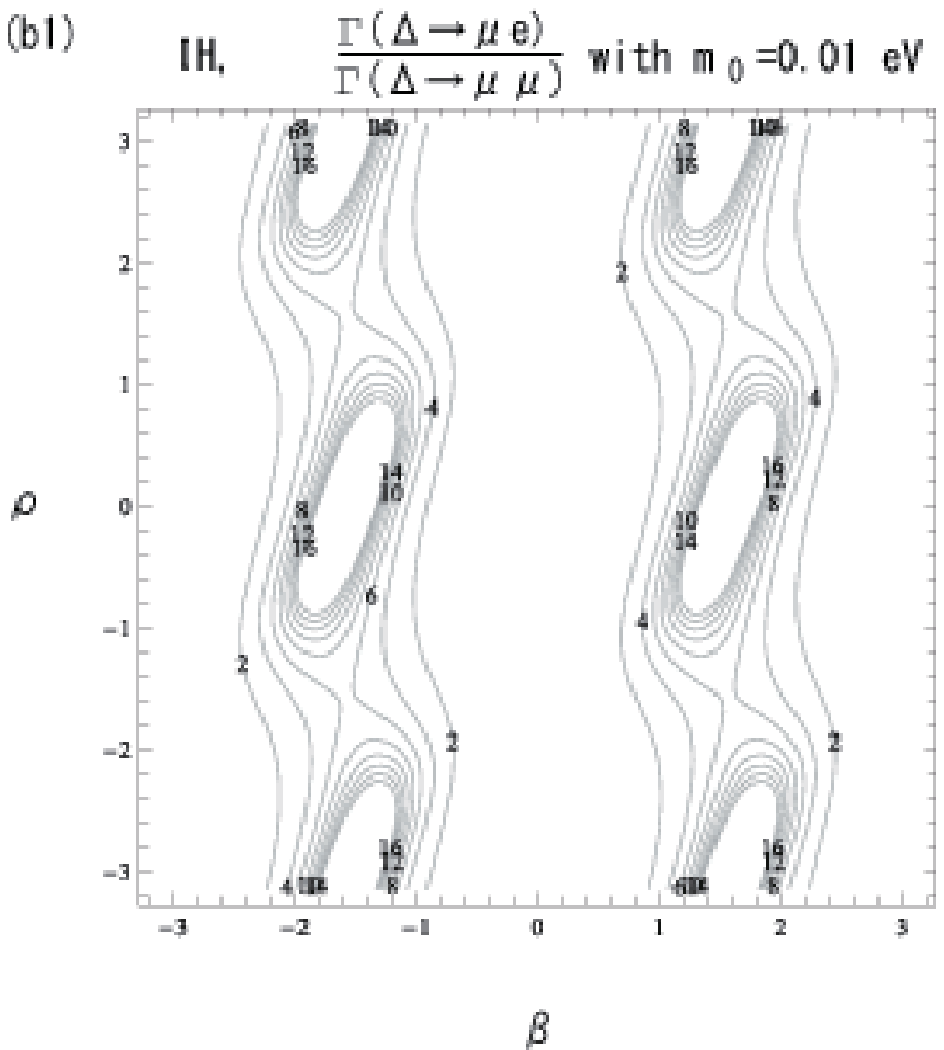}} } 
{\scalebox{0.75}{\includegraphics{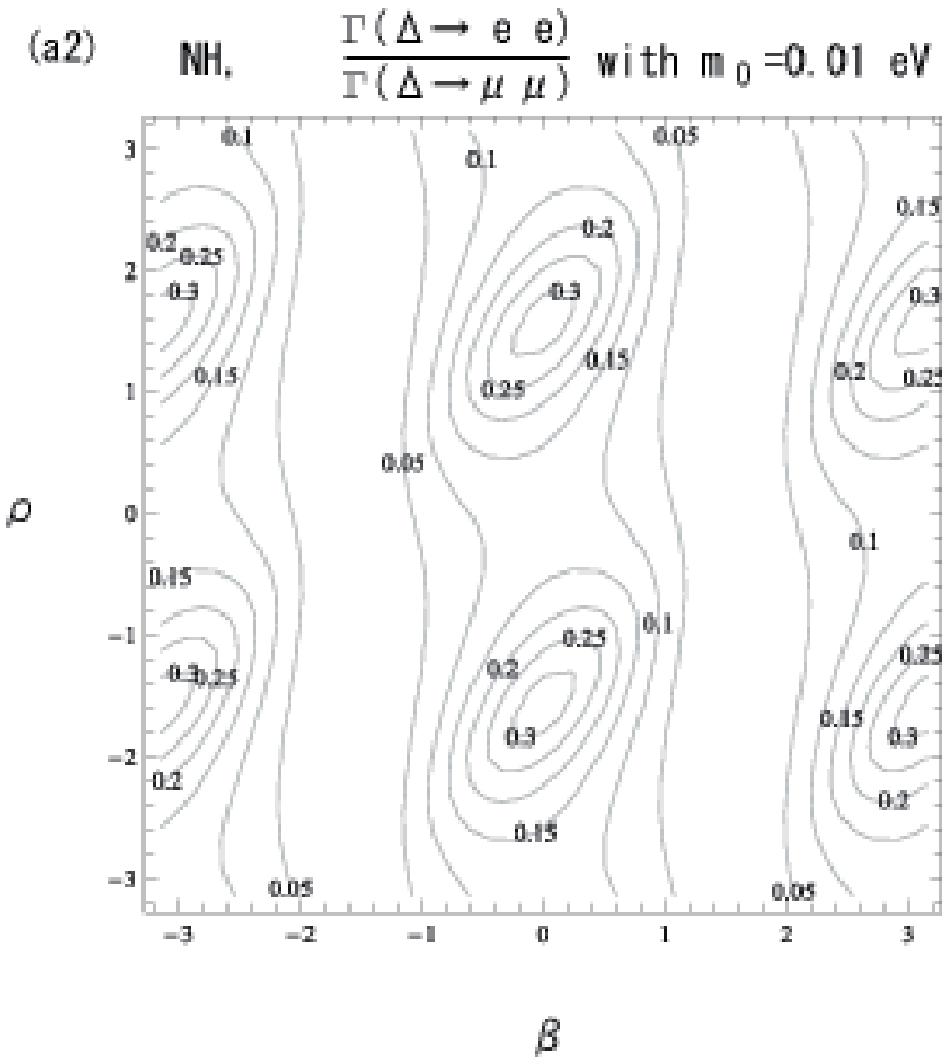}} }\ 
{\scalebox{0.75}{\includegraphics{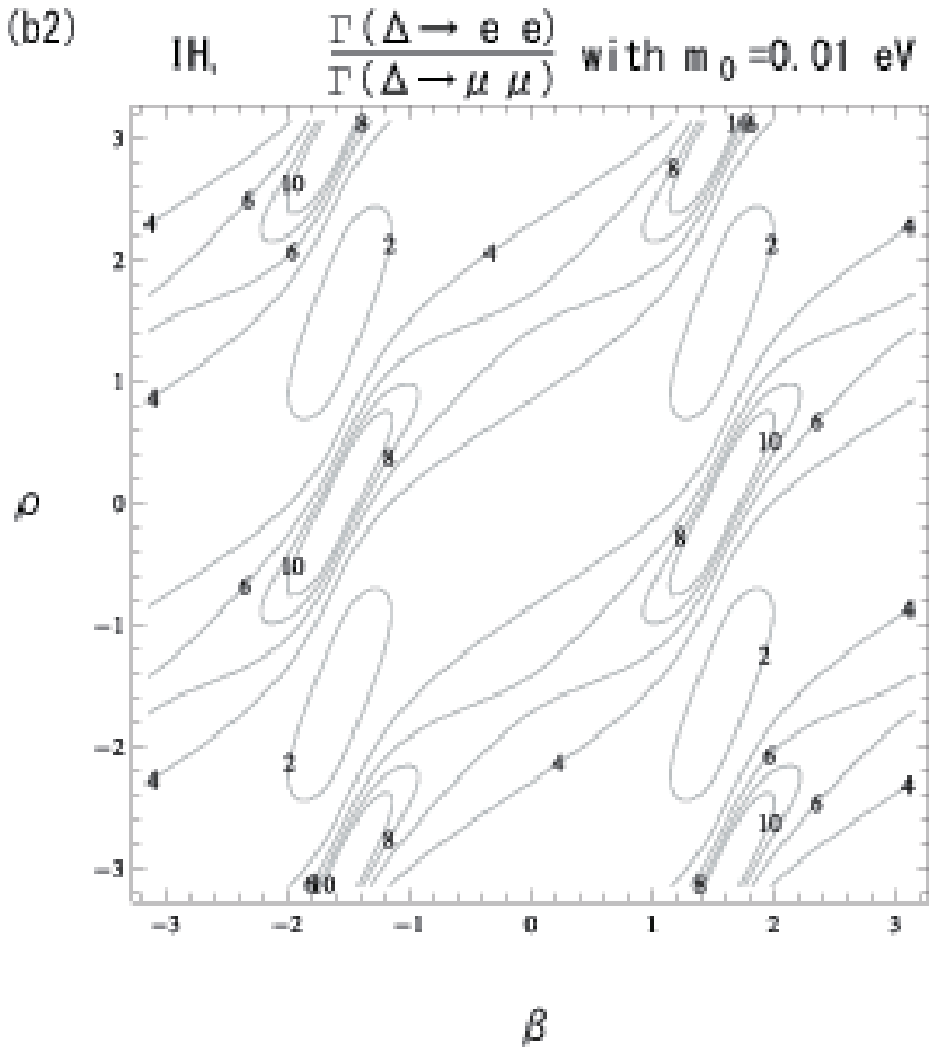}} }
\end{center}
\begin{quotation}
{\bf Fig.~7 }  Contour curves of  $\frac{\Gamma(\Delta^{--}\rightarrow \mu e)}{\Gamma(\Delta^{--}\rightarrow \mu\mu)}$ (panel 1)
and $\frac{\Gamma(\Delta^{--}\rightarrow e e)}{\Gamma(\Delta^{--}\rightarrow \mu\mu)}$ (panel 2) in the $\beta$ - $\rho$ plane  
 for the case of  $m_0=0.01$ eV with sin$^2\theta_{13}=0$.
Indices a and b indicate the normal and the inverse hierarchy cases, respectively. 
On condition that we measure both $\frac{\Gamma(\Delta^{--}\rightarrow \mu e)}{\Gamma(\Delta^{--}\rightarrow \mu\mu)}$ 
and $\frac{\Gamma(\Delta^{--}\rightarrow e e)}{\Gamma(\Delta^{--}\rightarrow \mu\mu)}$,  
we can determine values of the Majorana phase 
$\beta$ and $\rho$  from the intersections of the contour curves of (a1) and (a2) for the normal hierarchy case, 
and of (b1) and (b2) for the inverse hierarchy case, respectively.
\end{quotation}


\begin{thebibliography}{99}
\bibitem{Maki}
Z.~Maki, M.~Nakagawa, and S.~Sakata,
Prog.\ Theor.\ Phys.\  {\bf 28}, 870 (1962).
\bibitem{Fukuyama1}
K.~Matsuda, Y.~Koide, and T.~Fukuyama, Phys.\ Rev. {\bf D64}, 053015 (2001), arXiv:hep-ph/0010026;
K.~Matsuda, Y.~Koide, T.~Fukuyama, and H.~Nishiura, Phys.\ Rev. {\bf D65}, 033008 (2002), Erratum-ibid. {\bf D65}, 079904 (2002), arXiv:hep-ph/0108202;
T.~Fukuyama, N.~Okada, JHEP, 0211:011(2002), arXiv:hep-ph/0205066.
\bibitem{HTM}
W.~Konetschny and W.~Kummer, Phys.\ Lett.\  B {\bf 70}, 433 (1977);
J.~Schechter and J.~W.~F.~Valle, Phys.\ Rev.\  {\bf D22}, 2227 (1980);
T.~P.~Cheng and L.~F.~Li, Phys.\ Rev.\  {\bf D22}, 2860 (1980).
\bibitem{Fukuyama}
T.~Fukuyama and K.~Tsumura, arXiv:hep-ph/0809.5221.
\bibitem{Kotani}
M.~Doi, K.~Kotani, H.~Nishiura, K.~Okuda, and E.~Takasugi, Prog.\ Theor.\ Phys.\  {\bf 67}, 281 (1982).
See also M.~Doi, K.~Kotani, and H.~Nishiura, Prog.\ Theor.\ Phys.\  {\bf 118}, 1069 (2007). 
\bibitem{Han}
E.J.~Chun, K.Y.~Lee, and S.C.~Park, Phys.\ Lett.\ {\bf B566}, 142 (2003); 
A.G.~Akeroyd and M.~Aoki, Phys.\ Rev.\  {\bf D72}, 035011 (2005); 
T.~Han, B.~Mukhopandhyaya, Z.~Si, and K.~Wang, Phys.\ Rev.\  {\bf D76}, 075013 (2007); 
A.G.~Akeroyd, M.~Aoki, and H.~Sugiyama, Phys.\ Rev.\  {\bf D77}, 075010 (2008); 
J.~Garayoa and T.~Schwetz, JHEP, 0803, 009 (2008); 
M.~Kadastik, M.~Raidal, and L.~Rebane, Phys.\ Rev. {\bf D77}, 115023 (2008); 
P.F.~P$\acute{e}$rez {\it et al.}, Phys.\ Rev.\  {\bf D78}, 015018(2008); 
Del Aguila {\it et al.}, Nucl.\ Phys.\ {\bf B813}, 22 (2009). 
\bibitem{N-F}
H.~Nishiura and T.~Fukuyama, Phys.Rev. {\bf D80}, 017302 (2009).
\bibitem{Mu3eExp}
U.~Bellgardt {\it et al.}  (SINDRUM Collaboration), Nucl.\ Phys.\  {\bf B299}, 1 (1988).
\bibitem{Han2}
T.~Han, H.E.~Logan, B.~Mukhopandhyaya, and R.~Srikanth, Phys.\ Rev.\ {\bf D72}, 053007 (2005).
\bibitem{rhoHTM}
M.~Czakon, M.~Zralek, and J.~Gluza, Nucl.\ Phys.\  {\bf B573}, 57 (2000);
M.~Czakon, J.~Gluza, F.~Jegerlehner, and M.~Zralek, Eur.\ Phys.\ J.\  {\bf C13}, 275 (2000);
M.~C.~Chen and S.~Dawson, Phys.\ Rev.\  {\bf D70}, 015003 (2004);
M.~C.~Chen, S.~Dawson, and T.~Krupovnickas, Int.\ J.\ Mod.\ Phys.\  {\bf A21}, 4045 (2006).
\bibitem{tri}
P.~F.~Harrison, D.~H.~P$\acute{e}$rkins, and W.~G.~Scott, Phys.\ Lett. {\bf B530}, 167 (2002),  arXiv:hep-ph/0202074.
\bibitem{F-N}
T.~Fukuyama and H.~Nishiura, Proceeding of 1997 Shizuoka Workshop on Masses and Mixings of Quarks and Leptons, 19-21 (1997), arXiv:hep-ph/9702253.
\bibitem{PDG}
C.~Amsler {\it et al.}, Phys.\ Lett.\ {\bf B667},  1 (2008).
\bibitem{single}
A.~Osipowicz {\it et al.} (KATRIN Collaboration), arXiv:hep-ex/0109033.
\bibitem{MTFN}
H.V.~Klapdor-Kleingrothaus, H.~Pas, and A.Y.~Smirnov, Phys.\ Rev.\ {\bf D63}, 073005 (2001);

K.~Matsuda, N.~Takeda, T.~Fukuyama, and H.~Nishiura, Phys.\ Rev.\ {\bf D64}, 013001 (2001):
S. M.~Bilenky, S.~Pascoli, S.T.~Petcov, Phys.\ Rev.\ {\bf D64}, 053010 (2001);
For the recent review see, for instance,
R.N.~Mohapatra {\it et al.}, Theory of neutrinos: A White paper in Rept.\ Prog.\ Phys.\ {\bf 70}, 1757-1867 (2007), 
arXiv:hep-ph/0510213. 
\bibitem{betadecay} 
There are many on-going experiments, which are found in
F.~Avignone, Nucl.\ Phys.\ Proc.\ Suppl.\ {\bf 143}, 233 (2005);
See also NEXT Collaboration, arXiv:hep-ex/0907.4054; NEMO Collaboration, arXiv:hep-ex/0901.2720.
\bibitem{longbase}
Y.~Itow {\it et al.}(T2K Collaboration), arXiv:hep-ex/0106019;
F.~Ardellier {\it et al.} (Double Chooz Collaboration), arXiv:hep-ex/0606025.
\end{thebibliography}
\end{document}